\documentclass[floatfix,aps,prl,amsmath,twocolumn,amssymb,titlepage,superscriptaddress]{revtex4-1}
\usepackage{amsthm,amsmath,amsfonts,dsfont} 
\usepackage{graphicx}   
\usepackage{verbatim}   
\usepackage{subfigure}  
\usepackage{hyperref}   
\usepackage[ngerman,english]{babel}
\usepackage[usenames, dvipsnames]{xcolor}
\usepackage{tikz}
\usepackage{pgfplots}
\usepgfplotslibrary{colorbrewer}
\usepackage{tikz-3dplot}
\usepackage[percent]{overpic}
\usetikzlibrary{plotmarks,arrows.meta,calc,decorations.markings,math,arrows.meta,decorations.pathreplacing,decorations.pathmorphing}
\def\be{\begin{equation}}
\def\ee{\end{equation}}
\usepackage{braket}
\usepackage{makecell}
\usetikzlibrary{backgrounds,fit,decorations.pathreplacing}

\newcommand{\ketbra}[1]{\mathopen{|}#1\mathclose{\rangle}\hspace{-0.25em}\mathopen{\langle}#1\mathclose{|}}
\newcommand{\ketbrap}[2]{\mathopen{|}#1\mathclose{\rangle}\hspace{-0.25em}\mathopen{\langle}#2\mathclose{|}}

\begin{document}

\title{Quantum Zeno-based Detection and State Engineering of Ultracold Polar Molecules}

\author{Amit Jamadagni}
\email{amit.jamadagni@itp.uni-hannover.de}
\affiliation{Institut f\"ur Theoretische Physik, Leibniz Universit\"at Hannover, Appelstra{\ss}e 2, 30167 Hannover, Germany}
\author{Silke Ospelkaus}
\affiliation{Institut f\"ur Quantenoptik, Leibniz Universit\"at Hannover, Welfengarten 1, 30167 Hannover, Germany}
\author{Luis Santos}
\affiliation{Institut f\"ur Theoretische Physik, Leibniz Universit\"at Hannover, Appelstra{\ss}e 2, 30167 Hannover, Germany}
\author{Hendrik Weimer}
\affiliation{Institut f\"ur Theoretische Physik, Leibniz Universit\"at Hannover, Appelstra{\ss}e 2, 30167 Hannover, Germany}

\begin{abstract}
We present and analyze a toolbox for the controlled manipulation of
ultracold polar molecules, consisting of detection of molecules,
atom-molecule entanglement, and engineering of dissipative
dynamics. Our setup is based on fast chemical reactions between
molecules and atoms leading to a quantum Zeno-based collisional
blockade in the system. We demonstrate that the experimental
parameters for achieving high fidelities can be found using a
straightforward numerical optimization. We exemplify our approach for
a system comprised of NaK molecules and Na atoms and we discuss the
consequences of residual imperfections such as a finite strength of
the quantum Zeno blockade.

\end{abstract}
\maketitle

Ultracold polar molecules are very promising candidates for a wide
range of applications such as quantum simulation \cite{Moses2017} or
searching for physics beyond the Standard Model
\cite{Tarbutt2013}. However, the lack of optical control over the
internal states due to the large number of vibrational levels
currently presents a severe obstacle towards realizing these
applications. Here, we show that strong chemical reactions between
molecules and atoms enable a dissipative interaction mechanism that in
the quantum Zeno regime allows to leverage the optical properties of
atoms and achieve the same level of control for the molecules.

In current experiments, ultracold molecules are photo-associated from
laser cooled atoms \cite{Ni2008,DeMarco2019} with full control over
the final hyperfine state \cite{Ospelkaus2010a}. The resulting
molecules can be manipulated by external electric fields, microwave
driving, or optical potentials, e.g., enabling to load the molecules
into an optical lattice
\cite{Miranda2011,Reichsollner2017}. Crucially, ultracold molecules
can be subject to chemical reactions, both between themselves as well
as with their constituent atoms \cite{Ospelkaus2010,Guo2018}. For some
molecules such as NaK \cite{Park2015,Will2016,Seesselberg2018}, the
chemical stability with respect to these atom-molecule reactions
depends on the atom species involved \cite{Zuchowski2010}.

In this Letter, we present a complete toolbox for transferring the
optical control over atoms to the realm of polar molecules. Our
approach crucially relies on the presence of strong chemical reactions
between atoms and molecules. In the quantum Zeno regime
\cite{Zhu2014}, this dissipative atom-molecule interaction implements
a collisional blockade that allows for the efficient detection of
molecules, overcoming a longstanding challenge. In contrast to
previous proposals \cite{Covey2018}, our approach does not require
high-fidelity association and deassociation of the
molecules. Furthermore, we demonstrate how to extend our detection
method to the generation of atom-molecule entanglement and to the
controlled dissipation of rotational excitations of the molecules. We
dicuss the expected fidelities for experimentally realistic
parameters, as well as the role of residual imperfections such as an
imperfect quantum Zeno blockade.

In the following, we will first discuss the basic mechanism to implement a
detection scheme for the molecules. Later on, we will show how this
can be generalized towards the generation of entanglement between the
internal degrees of freedom of a molecule and an atom. Finally, we
will show how the entangling operation can also be employed towards
the controlled dissipation of rotational excitations of the molecule,
which can be readily used within laser cooling protocols.

\begin{figure}[t]
  \centering
  \includegraphics[width=0.95\linewidth]{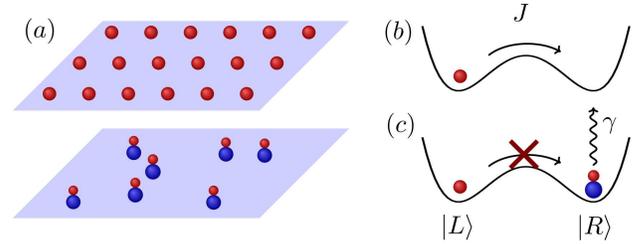}

  \caption{Quantum Zeno detection of molecules. (a) Bilayer setup of
    the system with the atoms being in a Mott insulator state in the
    upper layer separated from the molecules. (b) In the absence of a
    molecule in the neighboring site, atoms can transfer from
    $\ket{L}$ to $\ket{R}$. (c) The presence of a molecule inhibits the transfer to $\ket{R}$ due to a strong chemical reaction.
\label{fig1}}
\end{figure}
\emph{Detection of molecules.---} For the detection of the molecules,
we consider a bilayer setup of two two-dimensional optical lattice
potentials, see Fig.~\ref{fig1}. Different layers for the atoms and
molecules can be selected by magnetic and electric field gradients,
respectively, in combination with microwave fields. Hence, it is
possible to initially load all the Na atoms in one layer, while the
NaK molecules are located in the other layer. We assume that the
tunneling within the two lattice planes is strongly suppressed by the
lattice potentials, hence it is sufficient to consider a single
double-well setup. Additionally, we consider the atomic layer prepared
in a Mott insulator state with one atom per site. Owing to the
differences in the AC polarizability \cite{Holmgren2010,
  Neyenhuis2012}, as well as the fact that the atoms are much lighter
than the molecules, we limit the following analysis to the case where
only the atom moves with a hopping matrix element $J$, while the
molecule remains stationary throughout the process. The crucial
dissipative element is provided by a loss with a rate $\gamma$ induced
by the chemical reaction between Na atoms and NaK molecules
\cite{Idziaszek2010, Ospelkaus2010}.

The central idea behind the detection mechanism is to attempt a
Landau-Zener transfer of the atom to the site potentially occupied by
a molecule, e.g., by adiabatically reversing the magnetic field
gradient. If there is no molecule on the second site, the transfer
will succeed and the atom can be detected using standard quantum gas
microscope techniques \cite{Bakr2009, Sherson2010,
  Haller2015}. However, in the presence of molecule in the second
site, the strong chemical reaction creates a quantum Zeno blockade
\cite{Zhu2014} of the Landau-Zener process. Hence, detecting the
position of the atom after completion of the Landau-Zener sweep allows
to deduce the positions of the molecules.

\begin{figure}[t]
\centering
\includegraphics[trim=0.cm 0.4cm 0.cm 0.cm, width=\linewidth, clip]{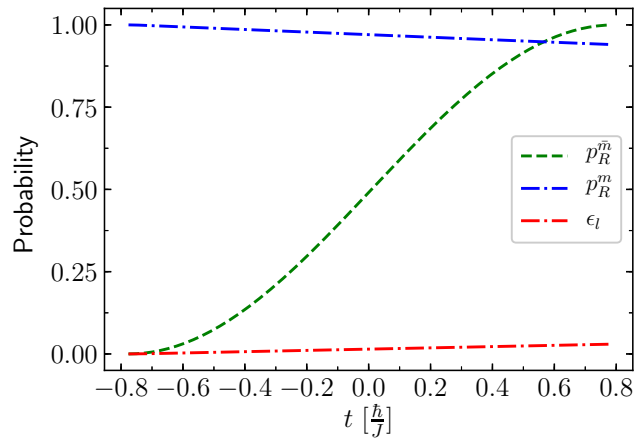}

\caption{Atom transfer probability for the atom in the presence ($p_R^m$) or absence ($p_R^{\bar{m}}$) of the molecule. Additionally, the loss $\varepsilon_l = 1-p_L^m-p_R^m$ is shown ($\gamma = 50\,J$).
\label{fig2}}
\end{figure}

To be specific, we model the detection in terms of two independent
Landau-Zener processes, with and without the presence of a molecule on
the second site. For simplicity, we assume half filling of the
molecular layer, i.e., equal probabilities for both processes, however
the subsequent optimization can also be tailored towards other
molecular densities. Then, the probability for a detection failure
(either false positive or false negative) is simply given by the
averaged error probability of the two processes. We model the Landau-Zener processes in terms of a non-Hermitian Hamiltonian
\begin{equation}
  H = \Delta(t) \sigma_z - J \sigma_x -\frac{i\gamma}{2}\ketbra{R},
\end{equation}
where we have used Pauli matrices $\sigma_\alpha$ in the basis of the
two atom positons $\ket{L}$ and $\ket{R}$, while $\Delta(t)$ describes
the influence of the time-dependent magnetic field gradient. The
dissipation $\gamma$ vanishes in the absence of a molecule on the
right site. The time-dependent term $\Delta(t)$ is taken to be an odd
polynomial in $t$ up to fifth order, i.e., satisfying the constraints
$\Delta(\pm\infty) = \pm \infty$ and $\Delta(0)=0$. Then, we can
minimize the total detection error $\varepsilon$ by optimizing the
coefficients of the polynomial as well as the initial time $t_i$,
while the final time $t_f$ is given by $t_f=-t_i$. As shown in
Fig.~\ref{fig2}, both Landau-Zener processes can be implemented with
high success rate, i.e., the total error $\varepsilon=0.03$ is low. We
note that this error can be even further reduced when starting from a
high fidelity Mott insulator for the atoms \cite{Bakr2011}, as
observing no atom in either site clearly signals the presence of a
molecule.

\begin{figure}[t]
\centering
\includegraphics[width=\linewidth]{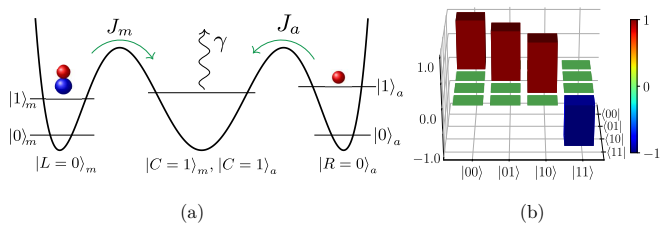}

\caption{Atom-molecule entanglement. (a) Three-well setup for the
  system with a central reaction site $\ket{C}$, while molecule and
  atom start in $\ket{L}$ and $\ket{R}$, respectively. (b) Matrix
  representation of the optimized process, realizing a CZ gate with
  high fidelity ($\gamma = 50\,J$).}
\label{fig6}
\end{figure}

\emph{Atom-molecule entanglement.---} As the next step, we extend the
Zeno-based detection scheme towards entanglement of the internal
states of atoms and molecules. The internal states $\ket{0}_{a/m}$ and
$\ket{1}_{a/m}$ refer to the rovibrational ground state and a
rotationally excited state for the molecule ($m$) and two hyperfine
states of the atom ($a$), respectively. To realize an entangling
operation, we consider a setup consisting of three two-dimensional
planes, of which we again consider a single three well unit, see
Fig.~\ref{fig6}a. We constrain the motion of the molecule (atom) to
the left (right) site and the center site, respectively. Additionally,
we assume that the motion takes only place within the $\ket{1}$ state
of both atom and molecule.

A possible scheme to engineer these constrains may be as follows. In
addition to the time-dependent magnetic or electric field gradient,
the central site $\ket{C}$ may be shifted by an energy difference
$\delta$ from that $\ket{L}$ and $\ket{R}$, such that direct hopping
is prevented. Sites $\ket{L}$ and $\ket{C}$~($\ket{R}$ and $\ket{C}$)
are connected along the $x$~($y$) direction. A first Raman pair (or rf
frequency) with frequency transfer $\Delta\omega_1=\Delta$, Rabi
frequency $\Omega_1$, and momentum transfer $\vec\Delta k_1=\Delta k_1
\vec e_x$, induces a molecular hopping rate $J_m=\hbar\Omega_1 \langle
1,L|_m e^{i\Delta k_1 x} |1,C\rangle_m$. Crucially, $\langle 1,R|_m
e^{i\Delta k_1 x} |1,C\rangle_m=0$ due to orthogonality of the
Wannier-like functions characterizing each lattice site. As a result,
molecules in the $|1\rangle_m$ state only hop from $\ket{L}$ to
$\ket{C}$. Similarly, using a second Raman pair with
$\Delta\omega_2=\Delta$, $\Omega_2$, and $\vec\Delta k_2=\Delta k_2
\vec e_y$, we may induce a hopping rate $J_a=\hbar\Omega_2 \langle
1,R|_a e^{i\Delta k_2 y} |1,C\rangle_a$, for atoms between $\ket{R}$
and $\ket{C}$ without any atomic hopping from $\ket{C}$ to $\ket{L}$.

The Hamiltonian of the three-well setup is of the form
\begin{align}
  H &= B \ketbra{1}_m + E_{HFS}\ketbra{1}_a + \Delta_m(t)\ketbra{L}_m  \nonumber\\
  &+ \Delta_a(t)\ketbra{R}_a -\frac{i\gamma}{2}\ketbra{C}_m\otimes \ketbra{C}_a \nonumber\\& - (J_m\ketbrap{1,L}{1,C}_m+ J_a\ketbrap{1,R}{1,C}_a + \text{H.c}),
\end{align}
where $B$ is the rotational constant of the molecule, $E_{HFS}$ is the
hyperfine splitting of the atom and $\Delta_{m/a}$ refers to the
electric and magnetic field gradients, respectively. For simplicity,
we assume the condition $\Delta_m(t)=-\Delta_a(t)\equiv \Delta(t)$,
although this choice is not essential for achieving high fidelity
entanglement.

The entangling operation we are realizing corresponds to a
Controlled-Z (CZ) quantum gate for the internal states. We restrict
the computational states to the molecule being in the motional state
$\ket{L}$ and the atom being in $\ket{R}$, i.e., any residual
occupation of the central site constitutes a gate error. Without loss
of generality, we set the tunnelling matrix elements to be identical,
i.e., $J_a=J_m\equiv J$. The CZ gate is implemented using a similar
Landau-Zener protocol as our detection scheme, however the constraints
on $\Delta(t)$ are different. Since the motional degrees of freedom
need to return to the initial value at the end of the gate, this
requires $\Delta(t)=\Delta(-t)$, therefore we take $\Delta(t)$ to be
an even polynomial in $t$ up to fourth order. Then, optimizing the
coefficients of the polynomial as well as the gate time result in a
close approximation to an exact CZ gate, see Fig.~\ref{fig6}b. To
capture the performance of the gate in a single number, we use the
fidelity $F = |\braket{\psi_+|\psi(t_f)}|^2$ to create the maximally
entangled state $\ket{\psi_+} = \frac{1}{\sqrt{2}}(\ket{0L_{m}0R_{a}}
+ \ket{1L_{m}1R_{a}})$, as this constitutes a worst case scenario for
a large class of error mechanisms \cite{Weimer2012}. For a dissipation
strength of $\gamma = 50\,J$, we can achieve a fidelity of $F=0.913$,
which approaches a regime that is sufficient for the digital quantum
simulation of important many-body spin models
\cite{Weimer2013a}. Finally, we would like to note that the presented
atom-molecule entanglement can be readily converted into
molecule-molecule entanglement using standard entanglement transfer
protocols \cite{Barenco1995}.

\begin{figure}[t]
\centering
\includegraphics[trim=0.cm 0.3cm 0.cm 0.1cm, width=\linewidth, clip]{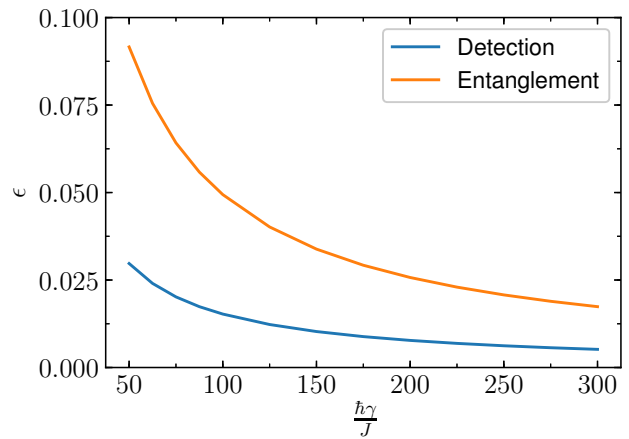}
\caption{Errors $\varepsilon$ for the detection and entangling process
  as a function of $\gamma/J$. Both curves follow a quantum Zeno
  profile proportional to $J/\gamma$.}
\label{fig4}
\end{figure}

In the previous analyses, we have assumed that the loss rate $\gamma =
50\,J$ is fixed with respect to the tunnelling matrix
element. However, by tuning the laser parameters for generating the
potential wells, it is possible to make this ratio highly
tunable. Therefore, it is interesting to investigate how our protocols
responds to such a change, it particular there might be practical
limitations on how low $J$ can become before other error mechanisms
such as thermal losses, buffer gas collisions, etc. kick
in. Figure~\ref{fig4} shows that both the detection and the
entanglement error $\varepsilon = 1-F$ both follow the expected
$1/\gamma$ quantum Zeno profile, with the detection error being
slightly below the entanglement error.

\emph{Dissipative quantum state engineering.---} Let us now extend the
mechanism behind the CZ gate towards the controlled dissipation of
rotational excitations. For polar molecules, such a mechanism cannot
reliably be realized using optical pumping due to the large number of
vibrational levels an optical excitation can decay into. Nevertheless,
such a controlled dissipation channel has immediate application for
the laser cooling of molecules \cite{Zhao2012}, as well as for the
dissipative engineering of tailored quantum many-body states
\cite{Diehl2008,Verstraete2009,Weimer2010}.

To realize this dissipative state engineering, we consider two
different strategies. The first strategy is to apply a pulsed scheme,
in which the CZ gate $U_{CZ}$ is combined with single qubit gates in
both the atom and the molecule, in close analogy to a digital
quantum simulator. We consider the gate sequence
\begin{equation}
  U = U_{CZ}R_x^{(a)}(\phi)U_H^{(m)}U_{CZ}U_H^{(m)}R_x^{(a)}(\phi),
\end{equation}
where $R_x(\phi) = \exp(i\phi\sigma_x)$ is the single qubit rotation
about the $x$ axis and $U_H$ is the Hadamard gate. After the gate
sequence, the molecule and the atom will be entangled. However, a
subsequent optical pumping of the atom according to $\ket{1} \to
\ket{0}$ realizes a discrete quantum master equation in Lindblad form according to
\begin{equation}
  \rho_m' = \rho_m  + \phi^2 \sigma_-^{(m)}\rho_m\sigma_+^{(m)} - \frac{\phi^2}{2}\left\{\sigma_+^{(m)}\sigma_-^{(m)},\rho_m\right\} + O(\phi^4).
\end{equation}
This dissipative dynamics can then be interspersed with coherent
driving of the rotational states of the molecules to achieve the
desired quantum state engineering.

\begin{figure}[t]
\centering
\includegraphics[trim=0.cm 0.325cm 0.cm .2cm, width=\linewidth, clip]{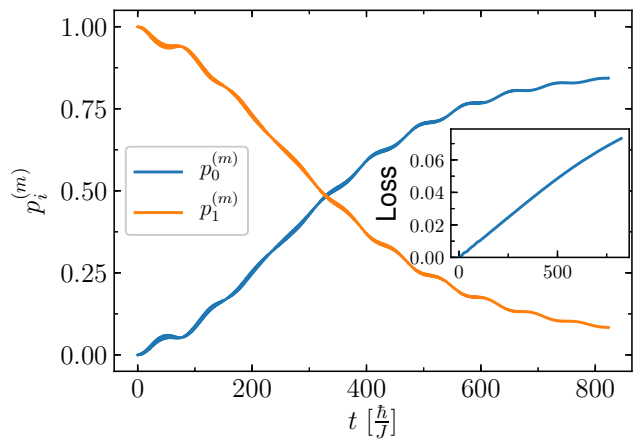}

\caption{Dissipation of a rotational excitation. The probability to
  find the molecule in the rotationally excited state $\ket{1}$,
  $p_1^{(m)}$ closely follows an exponential decay, while the ground
  state probability $p_0^{(m)}$ increases in a similar way. The inset
  shows that the overall loss from the chemical reaction remains low
  ($\gamma = 50\,J$).}
\label{fig8}
\end{figure}

The second strategy for this dissipative state engineering uses a
continuous driving of both the molecular and the atomic transitions
rather than a pulsed quantum gate sequence. This has two
advantages. On the one hand, the mechanism behind the CZ gate can be
understood to implement an effective $\sigma_z^{(m)}\sigma_z^{(a)}$
interaction, which cannot transfer energy from the molecule to the
atom. By going into the rotating frame of the driving, this is
converted into a $\sigma_x^{(m)}\sigma_x^{(a)}$ interaction, which
contains terms that lead to an energy transfer. The second advantage
is that the energy difference between the rotational constant $B$ and
the hyperfine splitting $E_{HFS}$ is too large to observe direct
energy transfer from the molecule to the atom. In the rotating frame,
however, this large energy difference is converted into the energy
difference involving the respective detunings and Rabi frequencies of
the driving. When the differences in the detunings and the Rabi
frequencies vanish, efficient energy transfer becomes possible under a
condition very similar to the Hartmann-Hahn driving in nuclear
magnetic resonance \cite{Hartmann1962}. In this case, the Hamiltonian
is given by
\begin{equation}
\begin{split}
  H = & -J(\ketbrap{1,L}{1,C}_{m} + \ketbrap{1,R}{1,C}_{a} + \text{H.c.}) \\
& + \Omega(\sigma_{x}^{(m)} + \sigma_{x}^{(a)}) + \delta(\sigma_{z}^{(m)} + \sigma_{z}^{(a)}) \\
& - \Delta\ketbra{L} - \Delta\ketbra{R},
\end{split}
\end{equation}
where $\Omega$ is the Rabi frequency and $\delta$ is the detuning used
both for the driving of the molecule and the atom. Addionally, we
consider optical pumping of the atom by including a quantum jump
operator $c_\kappa = \sqrt{\kappa} \sigma_-^{(a)}$ in the quantum
master equation describing the dynamics,
\begin{equation}
\frac{d}{dt}\rho = -\frac{i}{\hbar}[H,\rho] + \sum\limits_{i=\gamma,\kappa} \left(c_i\rho c_i^\dagger - \frac{1}{2}\left\{c_i^\dagger c_i, \rho\right\}\right).
\label{eq:master}
\end{equation}
  Here, the jump operator $c_\gamma$ describes the occurence of a
  chemical reation, which removes both the atom and the molecule from
  the system. We then optimize the set of parameters $(\Omega, \delta,
  \Delta, \kappa)$ such that an initial state of $\ket{1}_m\ket{0}_a$
  goes to $\ket{0}_m\ket{0}_a$. Figure~\ref{fig8} shows that this
  optimization leads to the initial rotational excitation being
  dissipated with high probability, while the loss resulting from
  chemical reaction remains low. We would like to note that this
  continuous dissipative process might be experimentally easier to
  realize than a pulsed sequence, however coming at the expense of
  requiring longer preparation times. We checked the robustness of our
  protocol with respect to an imperfect experimental implementation of
  the optimized parameters. Upon decreasing the atomic Rabi frequency
  by $10\%$ compared to its optimal value, we observe that the final
  probability to reach the $\ket{0}_m\ket{0}a$ state decreses by
  merely $2\%$. This underlines the robustness of our optimization
  protocol.

\emph{Experimental parameters.---} In the following, we will calculate
the loss coefficient $\gamma$ for the exothermic reaction $\text{NaK}
+\text{Na} \to \text{Na}_2 + \text{K}$. The $\text{NaK} +\text{Na}$
system has a van der Waals coefficient of $C_6 = 3592\,\text{a.u.}$
\cite{Zuchowski2013}, from which we can calculate the universal rate
coefficient $\beta$ of the chemical reaction \cite{Idziaszek2010},
which is given by
\begin{equation}
  \beta = \frac{2\hbar^{3/2} \Gamma(1/4)^2}{(2\mu^5 C_6)^{1/4}} = 1.47 \times 10^{-10}\,\text{cm}^3\,\text{s}^{-1},
\end{equation}
where $\mu$ is the reduced mass. In a potential well corresponding to
an oscillator length of $a_0 = 100\,\text{nm}$, this will result in a
reaction rate of $\gamma = 50\,\text{kHz}$. Compared to the typical
tunnelling matrix element of $J \sim 100\text{--}1000\,\text{Hz}$
\cite{Bloch2008}, dissipation rates $\gamma = 50\text{--}500\,J$ deep
in the quantum Zeno regime can be achieved.

In summary, we have presented a toolbox for the controlled
manipulation of ultracold polar molecules via dissipative interactions
with atoms. Our results pave the way for important experimental
advances in the detection, entanglement, and dissipative quantum state
preparation of molecular quantum many-body systems. While we have
exemplified our approach for NaK, it can also be readily applied to
other molecules, provided there is at least one exothermic reaction
with one of its constituent atoms. An interesting question for future
work is whether the dynamics leading to the formation of ultracold
collision complexes \cite{Mayle2012} can be harnessed for coherent
manipulation of molecules in a similar way.

\begin{acknowledgements}

  This work was funded by the Volkswagen Foundation and the DFG within
  SFB 1227 (DQ-mat, projects A03 and A04) and under Germany's
  Excellence Strategy EXC-2123-A2 (QuantumFrontiers).

  \end{acknowledgements}

\bibliographystyle{aip}
\bibliography{main.bbl}

\end{document}